\documentclass[pre,amssymb,amsmath,floats,twocolumn,showpacs,superscriptaddress]{revtex4}

\usepackage{epsfig}

\begin{document}

\title{Minimal models of weighted scale-free networks}

\author{S. N. Dorogovtsev}
\email{sdorogov@fis.ua.pt}
\affiliation{Departamento de F{\'\i}sica da Universidade de Aveiro, 3810-193 Aveiro, Portugal}
\affiliation{A. F. Ioffe Physico-Technical Institute, 194021
  St. Petersburg, Russia} 

\author{J. F. F. Mendes}
\email{jfmendes@fis.ua.pt}
\affiliation{Departamento de F{\'\i}sica da Universidade de Aveiro, 3810-193 Aveiro, Portugal}

\date{}

\begin{abstract}

We consider a 
class 
of simple, non-trivial models of evolving weighted scale-free networks. 
The network evolution in these models is determined by attachment of new vertices to ends of preferentially chosen weighted edges and by updating the weights of these edges. Resulting networks have scale-free distributions of the edge weight, of the vertex degree, and of the vertex strength. We discuss situations where this mechanism 
operates. 
Apart of stochastic models of weighted networks, we introduce a wide class of deterministic, scale-free, weighted graphs with the small-world effect.    
We show also how one can easily construct an equilibrium weighted network by using a generalization of the configuration model.  

\end{abstract}


\maketitle

\section{Introduction}\label{s-introduction} 

\begin{figure}[b]
\epsfxsize=60mm
\epsffile{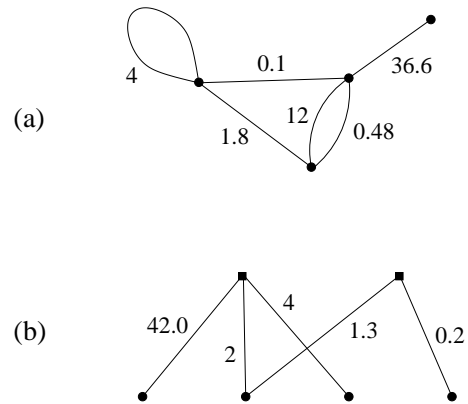}
\caption{
(a) A typical one-partite, weighted, undirected graph. 
(b) A bipartite weighted graph. 
}
\label{f1}
\end{figure}

All real-world networks are weighted. 
Let us 
explain this strong claim in more detail.  
Usual objects of interest in the science of networks are relatively simple nets where all edges have equal ``weights'' \cite{ab02,dm02,n03,dmbook03,pvbook04}. The elements of the adjacency matrices of these networks are ones and zeros. 
We stress that these simple graphs are only parodies of real networks where connections between vertices are not equal. Edge weights are introduced to describe this diversity.  In the simplest case, which we discuss here, weights are positive numbers \cite{yjbt01,n01,lm03,ztzh03,bbpv04,bbv04,bbv04a,lc03,flc04,mab04,lc04,r04,nr02,bbchs03,p04,yzwfw04,foc04} (see Fig.~\ref{f1}). 
So, the elements of the resulting adjacency matrices are zeros and positive numbers, $w_{ij}$. 
However, in principle, in more complex situation, edges may be described by a set of variables or operators. Note that an extra set of variables may be introduced to describe an individual properties of vertices (so called hidden variables \cite{gkk02,s02,cl02,ccrm02,bp03,mmk04}). 

Edge weights allow one to better describe reality. For example, in the simplest, unweighted version of a network of social contacts, each edge connects a pair of individuals who had one or two or more contacts. In a far more informative weighted network of social contacts, the weight of an edge shows the frequency of recent contacts between the corresponding pair of persons \cite{g73,g83}. 
Another example is a network of coauthorships, where edge weights show the number of joint papers of two coauthors \cite{lc03,flc04}. (This network may be treated as a one-mode projection of a bipartite graph.) Note that networks with multiple connections can be considered as weighted ones with integer edge weights.  

\begin{figure}
\epsfxsize=68mm
\epsffile{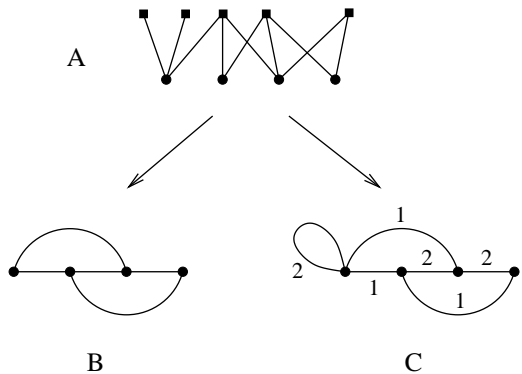}
\caption{
One-mode projections of a bipartite collaboration graph---graph~A. 
The circles and squares show the collaborators and the acts of collaboration respectively. 
Graph B is the unweighted one-mode projection of graph A. It is impossible to recover bipartite graph A by knowing unweighted graph B.
Graph C, which is the weighted one-mode projection of graph A, is more informative than graph B but, nevertheless, one still cannot recover graph A if the weighted projection C is known. Edge weights show the number of collaboration acts between the corresponding vertices.
}
\label{f2}
\end{figure}

Weighted networks provide an informative, usable picture of reality. 
Our first example is collaboration networks. Collaborations are adequately represented by bipartite collaboration graphs, where one kind of vertices is collaborators and the other one is acts of collaboration [see Fig.~\ref{f2}, graph~A]. Usually, empirical researchers consider one-mode projections of these graphs, where a pair of vertices-collaborators is interconnected by an edge if there has been at least one act of collaboration between them [see Fig.~\ref{f2}, graph B]. This net is more simple than the original bipartite graph, but an essential information is lost. In particular, one cannot recover the bipartite original by using this projection. The weighted one-mode projection shown in Fig.~\ref{f2}, graph C, is more informative than unweighted projection B but, nonetheless, is less informative than original bipartite graph A. One still cannot recover bipartite graph A if weighted projection C is known. 


The second example is energy landscape networks, where each vertex is a local minimum of the potential energy landscape, and an edge connects two minima with a saddle point between them \cite{sab01,d02}. An unweighted version of this network reflects only a general structure of connections in the configuration space of a system. In contrast, directed weighted graphs, where edge weights indicate transition rates provide basic necessary information about the relaxation and kinetics of a system. (One may also ascribe the energies of the potential minima to the vertices, and the energies of the saddle points to the edges.)  


The third example is various chemical reaction networks \cite{sm04} (e.g., networks of metabolic reactions \cite{akvob04}). In a one-partite version of these networks, weights show the chemical reaction weights. Networks of corporate ownerships are also directed weighted graphs (edge weights indicate the fractions of company's shares in hands of other companies).  
The links of spacial networks \cite{gn04} has a natural characteristic---their length. So, in this case an edge weight may be introduced as some function of the length of the corresponding edge.   

The simplest local characteristics of weighted networks are (i) the weight of an edge, $w_{ij}$, (ii) the degree of a vertex, and (iii) ``the strength'' of a vertex, 

\begin{equation}
s_i \equiv \sum_{j \in i} w_{ij}
\, ,
\label{e1}
\end{equation} 
Here the sum is over the nearest neighbors of the vertex $i$. (One can introduce directed weighted networks, where vertices have ``in-strength'' and ``out-strength''.) The distributions of these local characteristics in random networks are a weight distribution $Q(w)$, a degree distribution $P(k)$, and a strength distribution $R(s)$. 

A number of models, which provide complex, in particular, scale-free weighted networks, were proposed (see, e.g., Refs.~\cite{yjbt01,lc04,ztzh03,bbpv04,bbv04,bbv04a,foc04}). Obviously, one may easily construct a complex weighted network independently ascribing degrees to vertices and weights to edges (see the next section). For this, (i) build an unweighted network by any of existing algorithms and afterwords (ii) arrange weights of edges. These are trivial constructions. A much more interesting scale-free network model, where the evolutions of degrees and weights are coupled, was proposed in 
Refs.~\cite{bbpv04,bbv04,bbv04a}. 

In this network, in addition to a standard preferential attachment rule \cite{ba99,baj99} (see also \cite{p76}), a specific redistribution of edge weights was introduced. In more detail, (i) each new vertex is attached to a preferentially selected existing vertex (the probability of the attachment is proportional to the vertex strength), and (ii) weights of connections of the latter vertex are updated in a 
specific way depending on the strength of the vertex. 
That is, in this model, a preferential attachment to a vertex induces changes in weights of its edges. 
This evolution results in (a) a power-law weight distribution, (b) a linear dependence of the strength of a vertex on its degree, and (c) power-law strength and degree distributions. 
Note that the specific form of rule (ii) is necessary to obtain scale-free networks. 

In the present paper we show that quite similar results [(a), (b), and (c)] can be obtained by using a more simple construction. Actually we use the same idea as in our network evolving due to attachment to the ends (or to an end) of a randomly chosen edge 
\cite{dms01}. In the present network, 
\begin{list}{}{} 

\item[(i)] 
the weight of a preferentially chosen edge is increased by a constant, and 

\item[(ii)] 
its end receives a new connection of a unit weight. 

\end{list}
One can see that this model is so simple, that it may be considered as a minimal one. 

The paper is organized as follows. In Sec.~\ref{s-equilibrium} we, for comparison, demonstrate a ``trivial'' construction of an equilibrium weighted network, which is a direct generalization of the configuration model. In Sec.~\ref{s-model} we define our stochastic model and present its solution. In Sec.~\ref{s-how_to_choose} we explain how one can choose edges at random or accounting for their weights. In Sec.~\ref{s-generalizations} we describe simple generalizations of the model. In Sec.~\ref{s-deterministic} we introduce deterministic scale-free weighted networks. In Sec.~\ref{s-applications} we discuss possible applications of our models.


\section{Equilibrium weighted networks}\label{s-equilibrium}

As a starting point, we consider a ``trivial'' equilibrium, weighted network---the simplest generalization of the configuration model of a random network. 
The standard configuration model is, in simple terms, a maximally random graph with a given degree distribution, $P(k)$, see Ref.~\cite{b80}, where the model was introduced in this form, and also Refs.~\cite{bbk72,bc78,w81} with very similar constructions. In addition, we assume that the weights of the edges in the random graph ensemble of the configuration model are random, independent, and distributed according to a given weight distribution $Q(w)$. The question is: what is the resulting distribution of the vertex strength, $R(s)$, in this network? 

For brevity, in this section, we consider only networks with integer edge weights. However, the final result for the asymptotics of distributions is also valid without this assumption.

Actually, we must calculate the distribution of the sum of independent equally distributed random variables. 
The resulting relation is especially simple in a $Z$-transformed form. Let us introduce the $Z$-transforms (generating functions) of the distributions $P(k)$, $Q(w)$, and $R(s)$: 
\begin{eqnarray}
& & 
\Phi(z) \equiv \sum_k z^k P(k)
\nonumber
\\[5pt]
& &
\Sigma(z) \equiv \sum_w z^w Q(w)
\nonumber
\\[5pt]
& &
\Psi(z) \equiv \sum_s z^s R(s)
\, . 
\label{e2}
\end{eqnarray} 
Then, accounting for the definition of the strength (\ref{e1}) and for the mutual independence of edge weights (and vertex degrees), we immediately have 


\begin{equation}
\Psi(z) = \sum_k P(k)\Sigma^k(z) = \Phi(\Sigma(z))
\, .
\label{e3}
\end{equation} 
Here we used a standard relation for the $Z$-transform of the distribution of the sum of independent random variables (see, e.g., Ref.~\cite{nsw01}). 

The asymptotic behavior of the original distribution at large values of a random variable is related to the analytical structure of the corresponding $Z$-transformation near $z=1$. For example, for a power-law original, we have the following correspondence: 
\begin{eqnarray}
& & 
\!\!\!\!\!P(k\gg1)\sim k^{-\gamma_k} \ \Longleftrightarrow
\nonumber
\\[5pt] 
\!\!\!\!\!\!\Psi(z\ \text{near}\ 1) &\cong& 
1 + \sum_{i>0}a_i (1-z)^i + b(1-z)^{\gamma_k-1}
\, . 
\label{e4}
\end{eqnarray} 
On the other hand, the $Z$-transformation of a rapidly decreasing distribution (i.e., with the finite moments), is analytical at $z=1$. 
So, accounting for correspondence (\ref{e4}), relation (\ref{e3}) leads to the following conclusions for this network:  
 
\begin{list}{}{} 

\item[(i)]
if both the distributions $P(k)$ and $Q(w)$ are rapidly decreasing, then the strength distribution $R(s)$ is rapidly decreasing too;  

\item[(ii)]
if $P(k)$ is rapidly decreasing, and $Q(w)$ is, e.g., a power law, $Q(w) \sim w^{-\gamma_w}$, then the resulting strength distribution is also power-law, with the same exponent, $R(s) \sim s^{-\gamma_w}$, i.e., $\gamma_s=\gamma_w$;  

\item[(iii)]
if $P(k)$ is, e.g., scale-free, $P(k)\sim k^{-\gamma_k}$, and $Q(w)$ rapidly decreasing (i.e., $\Sigma(z)$ analytic at $z=1$), then $R(s) \sim s^{-\gamma_k}$, i.e., $\gamma_s=\gamma_k$; 

\item[(iv)]
if both $P(k)$ and $Q(w)$ are scale-free, then the resulting distribution $R(s)$ is also scale-free with exponent $\gamma_s = \min(\gamma_k,\gamma_w)$. 

\end{list}
These are important, although particular, cases. 
However,  
relation (\ref{e3}) allows one to easily obtain, after an inverse $Z$-transform, a strength distribution for the weighted graph with arbitrary degree and weight distributions.  

Note that, also, one may consider a more ``serious'', ``non-trivial'' generalization of the configuration model. These are labeled random graphs with a given sequence of ``generalized degrees''. The generalized degree of a vertex is a complete set of numbers, where each number shows how many edges of a given weight are attached to this vertex. 
This is quite similar to the constructions of graphs with hidden color of Refs.~\cite{s03}. 


\section{Attachment to weighted edges}\label{s-model}

Let us introduce the model of a growing scale-free network in more precise terms than in the Introduction. We first discuss the simplest case. 

\begin{figure}
\epsfxsize=60mm
\epsffile{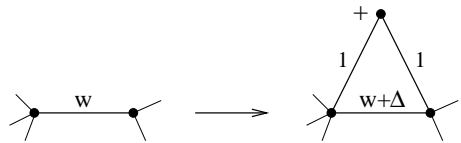}
\caption{
Schematic view of the network growth. At each time step the weight of a preferentially chosen edge is increased by a constant $\Delta$, and a new vertex is attached to the ends of this edge. 
}
\label{f3}
\end{figure}

We assume that the growth starts from an arbitrary configuration of vertices and edges, e.g., from a single edge of weight $1$. At each successive time step,  

\begin{list}{}{} 

\item[(i)]
choose an edge with probability proportional to its weight and increase this weight by a constant $\Delta \geq 0$, 

\item[(ii)]
attach a new vertex to both the ends of this edge by edges of weight $1$  

\end{list}
(see Fig.~\ref{f3}).


In the particular case of $\Delta=0$, this network is reduced to our model of Ref.~\cite{dms01}. 
Rule (i) may be interpreted in the following way. Suppose that $\Delta$ and the weights of the edges are integer numbers. In this case, an edge of weight $w$ may be treated as a $w$-multiple edge. If we choose each ``elementary'' edge in this network with multiple edges at random, with equal probability, we just arrive at the proportional preferential choice of edges in the corresponding weighted network. 

Rule (ii) may be easily modified: attach a new vertex to one of the ends of the selected edge and not to the both of them---see below. In the present form, rule (ii) results in the following rigid coupling of the degree $k_i$ of a vertex and its strength $s_i$: 

\begin{equation}
s_i = k_i(1+\Delta) - 2\Delta
\, .
\label{e5}
\end{equation} 
Indeed, in this model, each attachment to a vertex increases its degree by $1$ and its strength by $1+\Delta$ ($1$---due to the link of weight $1$ to a new vertex and $\Delta$ due to the modification of the selected edge). Taking into account that new vertices have degree and strength equal to $2$ results in relation (\ref{e5}). 
So the degree and strength distributions have quite similar asymptotic behaviors. 

We will show below that the weight distribution, the strength distribution, and the degree distribution of this network have power-law asymptotics: 

\begin{equation}
Q(w) \sim w^{-\gamma_w}, \ \ \ 
R(s) \sim s^{-\gamma_s}, \ \ \
P(k) \sim k^{-\gamma_k}
\label{e5a}
\end{equation} 
with exponents

\begin{eqnarray}
& & 
\gamma_w = 1 + \frac{2+\Delta}{\Delta} = 2 + \frac{2}{\Delta}  
\, , 
\label{e5b}
\\[5pt] 
& &
\gamma_s = \gamma_k = 1 + \frac{2+\Delta}{1+\Delta} = 2 + \frac{1}{1+\Delta}
\, . 
\label{e5c}
\end{eqnarray} 
Formula (\ref{e5b}) is valid at $\Delta>0$, and formula (\ref{e5c}) is valid at $\Delta \geq 0$. 


Note that formulas (\ref{e5b}), (\ref{e5c}) give $\gamma_s=\gamma_k<\gamma_w$. 
Recall the relation $\gamma_s=\text{min}(\gamma_k,\gamma_w)$, which we have obtained for scale-free equilibrium networks (the ``configuration model of a weighted network''). So, formulas (\ref{e5b}), (\ref{e5c}) satisfy that relation. 
One can easily check that the same is valid for the other scale-free weighted networks in this paper [see formulas (\ref{e14}) and (\ref{e15}), (\ref{e19}) and (\ref{e20}), (\ref{e21}) and (\ref{e22})]. 

Note that the evolution of edge weights in this model is quite similar to that in the Simon model \cite{s55}. One must take into account that in our network the total weight of edges growth proportionally to $t$: $W(t) \cong t(2+\Delta)$, which is asymptotic expression at large $t$. 
This gives the following evolution equation for the mean number $\overline{N}(w,t)$ of edges of weight $w$ at time $t$: 

\begin{eqnarray}
& & 
\overline{N}(w,t+1) = \overline{N}(w,t) + 2\delta_{w,1} 
\nonumber
\\[5pt] 
& &
+ 
\frac{w-\Delta}{t(2+\Delta)}\overline{N}(w-\Delta,t) - \frac{w}{t(2+\Delta)}\overline{N}(w,t)
\, . 
\label{e6}
\end{eqnarray} 
(For similar equations for degree distributions in growing scale-free networks, see Refs.~\cite{krl00} and \cite{dms00}.) 
At each time step two new edges of weight $1$ emerge. This produces the term $2\delta_{w,1}$ on the right-hand side of the equation. The third term describes increase in the number of edges of weight $w$ due to the modification of edges of weight $w-\Delta$. The last term on the right-hand side describes the reduction of the number of edges of weight $w$ due to their modification. 

We pass to the continuum $t$ limit and assume a stationary form of the weight distribution,
$\overline{N}(w,t) = 2tQ(w,t) \to 2tQ(w)$. 
Then, passing to the continuum limit of weight allows us to obtain a power-law form of the weight distribution $Q(w) \sim w^{-\gamma_w}$ 
with the $\gamma_w$ exponent given by formula (\ref{e5b}). 


The exact solution of the stationary limit of Eq.~(\ref{e8}),  

\begin{equation}
Q(w) \propto B(w/\Delta,\gamma_w)
\, , 
\label{e8}
\end{equation} 
is quite similar to the solution of the Simon model (we have omitted a normalization factor on the right-hand side of this formula). Here $B(\ ,\ )$ is the beta-function. As is natural, at large $w$, this expression has a power-law form with the same exponent $\gamma_w$ given by relation (\ref{e5b}). 

The total strength of the vertices in the network is $S(t)=2W(t) \cong 2t(2+\Delta)$. 
[The total degree of the network evolves in the following way: 
$K(t) \cong 4t$.] 
The attachment of a new vertex to the end vertices of an edge increases their strengths by $1+\Delta$. Furthermore, the rules of the model actually result in preferential (proportional) attachment to vertices of higher strength.
So, the evolution equation for the mean number $\overline{N}(s,t)$ of vertices of strength $s$ at time $t$ looks as follows: 

\begin{eqnarray}
& & 
\!\!\!\!\!\!\!\!\!\!\!\!\!\!\!\!\!
\overline{N}(s,t+1) = \overline{N}(s,t) + \delta_{s,2} 
\nonumber
\\[5pt] 
& &
\!\!\!\!\!\!\!\!\!\!\!\!\!\!\!\!\!
+ 
2\frac{s\!-\!(1\!+\!\Delta)}{2t(2+\Delta)}\overline{N}(s\!-\!(1\!+\!\Delta),t) - 2\frac{s}{2t(2+\Delta)}\overline{N}(s,t)
. 
\label{e9}
\end{eqnarray} 
The factors $2$ of the third and the fourth terms on the right-hand side of this equation are due to the attachment of each new vertex to the two ends of a selected edge. 
The resulting stationary strength distribution $R(s)$, $\overline{N}(s,t) = tR(s,t) \to R(s)$, is of the form 

\begin{equation}
R(s) \propto B(s/(1\!+\!\Delta),\gamma_s)
\, 
\label{e10}
\end{equation} 
with $\gamma_s$ given by formula (\ref{e5c}).   
Consequently, due to linear relation (\ref{e5}) between strength and degree, the resulting degree distribution is 
\begin{equation}
P(k) \sim B(k,\gamma_k)
\, 
\label{e12}
\end{equation} 
with $\gamma_k$ given by formula (\ref{e5c}). 
One can see that the resulting expressions of exponents, Eqs.~(\ref{e5a}) and (\ref{e5b}), 
are similar to those obtained for the model of Refs.~\cite{bbpv04,bbv04,bbv04a}. The ranges of variation with the parameter of a problem---$(2,\infty)$ for $\gamma_w$ and $(3,\infty)$ for $\gamma_s$ and $\gamma_k$---are the same in both these networks.  

By construction, our networks have numerous loops of length three, which indicates high clustering (see Ref.~\cite{dms01}). However, one should be careful with this claim. 
There are two distinct integrated clustering characteristics. 
One can show that the average clustering (the average clustering coefficient of a vertex) in this case is finite in the infinite network limit. 
On the other hand, the clustering coefficient (in simple terms, the density of triangles in the network) approaches zero in the infinite network.

\section{How to choose an edge at random}\label{s-how_to_choose}

A meticulous reader may say: ``Your model, as it was formulated, is indeed looks more simple than that of Refs.~\cite{bbpv04,bbv04,bbv04a}. However, your model is based on the selection of connections, which may be, computationally, more complex task than the selection of vertices. This may depreciate this model.'' 

To convince this reader, we indicate two easy (and quick) ways to select connections. 
(i) Make a list of edges (this list is sufficiently short in sparse networks) and choose from it. (ii) Choose edges, starting the procedure from random vertices. Below we describe the second way in more detail.    

Let us first demonstrate how one can select at random edges in a network. The simple procedure looks as follows:

\begin{list}{}{} 

\item[(i)]
choose a random vertex and then 

\item[(ii)]
subsequently choose each of its edges with some sufficiently small probability $p$,  


\end{list} 
then, repeat.  
Here, the probability $p$ must be small: $p < 1/k_{\text{max}}$, or at least 
$p \ll 1/\overline{k}$, where $k_{\text{max}}$ is the maximal degree of a vertex in the network, and $\overline{k}$ is the mean degree. 
This condition is necessary to subsequently select single edges but not bunches of edges---``hedgehogs''.  
(Note that, unlike the above procedure, if we simply select a random edge of a random vertex, we will not choose a random edge.)    

Similarly, in a weighted network, one can choose an edge with probability proportional to its weight. For this, 

\begin{list}{}{} 

\item[(i)]
Choose a random vertex and then 

\item[(ii)]
subsequently choose each of its (weighted) edges with some sufficiently small probability proportional to its weight.  


\end{list} 
then, repeat.  
Rule (ii) means that, after a vertex (say vertex $i$) is selected, each of its weighted edges is selected with probability $p w_{ij}/s_i$. 
Here, $s_i$ is the strength of this vertex [see definition (\ref{e1})] and $w_{ij}$ is the strength of an edge attached to this vertex. $p$ is an arbitrary parameter, so small that there is a small chance to subsequently select two or more edges attached to one vertex (see the above procedure).

\section{Simple generalizations}\label{s-generalizations}

\begin{figure}
\epsfxsize=60mm
\epsffile{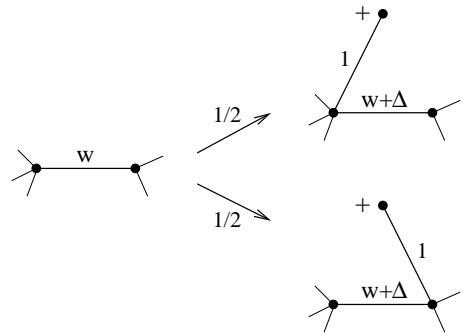}
\caption{
The scheme of the growth process for the modified network. At each time step the weight of a preferentially chosen edge is increased by a constant $\Delta$, and a new vertex is attached to any one of the ends of this edge. 
}
\label{f4}
\end{figure}

As we promised, let us consider the variation of the model with attachment to one end vertex of a selected edge.   
At each successive time step,  

\begin{list}{}{} 

\item[(i)]
choose an edge with probability proportional to its weight and increase this weight by a constant $\Delta \geq 0$, 

\item[(ii)]
attach a new vertex to one the ends of this edge by edges of weight $1$  

\end{list}
(see Fig.~\ref{f4}). 


The total weight, the total degree, and the total strength of the vertices of this network grow in the following way: $W(t) \cong t(1+\Delta)$, $K(t) \cong 2t$, $S(t) \cong 2t(1+\Delta)$, respectively. 
We will show that the network has power-law distributions (\ref{e5a}) with exponents 

\begin{eqnarray}
& & 
\gamma_w = 1 + \frac{1+\Delta}{\Delta} = 2 + \frac{1}{\Delta}  
\, , 
\label{e14}
\\[5pt] 
& &
\gamma_s = \gamma_k = 1 + 2\,\frac{1+\Delta}{1+2\Delta} = 2 + \frac{1}{1+2\Delta}
\, . 
\label{e15}
\end{eqnarray} 

Instead of rigid coupling (\ref{e5}) between the degree and strength of an individual vertex in the network of Sec.~\ref{s-model}, now we have only the asymptotic relation between the mean values of the strength and degree of an individual vertex: 
\begin{equation}
\overline{s}_i \cong (1 + 2\Delta)\overline{k}_i
\, . 
\label{e16}
\end{equation} 
This relation is valid for highly connected vertices (with large degrees). 
It shows that $\gamma_s = \gamma_k$ in this network.  

Asymptotic equality (\ref{e16}) follows from the following simple considerations. 
At each modification of vertex $i$, its degree and strength are modified in the following way: 
(i)~with probability $1/2$, $k_i \to k_i+1$ and $s_i \to s_i+\Delta+1$, and 
(ii) with probability $1/2$, $k_i \to k_i$ and $s_i \to s_i+\Delta$. This directly leads to relation (\ref{e16}). 

One can see that for this network, the equation for the mean number of edges of weight $w$ at time $t$ is

\begin{eqnarray}
& & 
\!\!\!\!\!\!\!\!\!\!
\overline{N}(w,t+1) = \overline{N}(w,t) + \delta_{w,1} 
\nonumber
\\[5pt] 
& &
\!\!\!\!\!\!\!\!\!\!
+ 
\frac{w-\Delta}{t(1+\Delta)}\overline{N}(w-\Delta,t) - \frac{w}{t(1+\Delta)}\overline{N}(w,t)
\, 
\label{e17}
\end{eqnarray} 
[compare with Eq.~(\ref{e6})]. This immediately leads to formula (\ref{e14}) for the $\gamma_w$ exponent of the weight distribution. 

For the mean number of vertices of strength $s$ in this network, we have the following equation: 

\begin{eqnarray}
& & 
\!\!\!\!\!\!\!\!\!\!
\overline{N}(s,t+1) = \overline{N}(s,t) + \delta_{s,1} 
\nonumber
\\[5pt] 
& &
\!\!\!\!\!\!\!\!\!\!
+
\left[\frac{s-(1\!+\!\Delta)}{2t(1+\Delta)}\overline{N}(s-(1\!+\!\Delta),t) - \frac{s}{2t(1+\Delta)}\overline{N}(s,t)\right]
\nonumber
\\[5pt] 
& &
\!\!\!\!\!\!\!\!\!\!
+ 
\left[\frac{s-\Delta}{2t(1+\Delta)}\overline{N}(s-\Delta,t) - \frac{s}{2t(1+\Delta)}\overline{N}(s,t)\right]
\label{e18}
\end{eqnarray} 
[compare with Eq.~(\ref{e9})]. This, together with relation (\ref{e16}) leads to formula (\ref{e15}) for the $\gamma_s$ exponent of the strength distribution and the $\gamma_k$ exponent of the degree distribution. 
One can see that the only difference between formulas (\ref{e5b}), (\ref{e5c}) and (\ref{e14}), (\ref{e15}) is that the parameter $\Delta$ in the results of Sec.~\ref{s-model} is now substituted by $2\delta$. 

In the same way as above, one can consider a combination of the models that we have already discussed. 
For example, the following combination of the evolution channels may be introduced. At each time step: (i) Choose preferentially an edge and increase its weight. (ii) (a) With probability $p_2$, attach a new vertex to both the ends of this edges; (b) with probability $p_1$, attach a new vertex to one of the end vertices of this edge; (c) with probability $p_0$, do not add a new vertex. 
Here, $p_0+p_1+p_2=1$.
One can also add the processes of linking of existing vertices, rewiring of connections, and introduce variation of $\Delta$.    

\begin{figure}
\epsfxsize=62mm
\epsffile{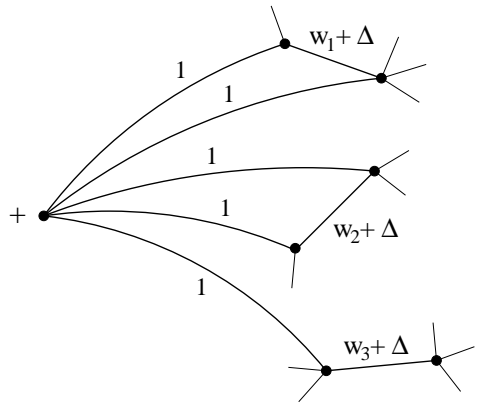}
\caption{
The growth process with attachment of a new vertex to several weighted edges.  
}
\label{f5}
\end{figure}

Up to now we discussed the proportional preferential choice of edges, that is, edges were chosen with probability proportional to their weights. 
In general, one may choose edges with probability proportional to some function $f(w)$ of their weight (preference function). 
Furthermore, one may consider an inhomogeneous situation, where a preference function depends on an edge $f_{ij}(w)$. Here the preference functions $f_{ij}(w)$ (and their parameters) may be randomly distributed similarly to what was considered for more traditional preferential attachments to vertices in inhomogeneous networks (see Refs.~\cite{bb01a,bb01b}). 
$\Delta$ also may be made random variable. 
One may also, at each time step, attach a new vertex to ends of several edges but not to a single edge as above (see Fig.~\ref{f5}). By using these generalizations, we can easily obtain various complex weighted network architectures with fat-tailed and rapidly decreasing distributions, with condensation, gelation, etc., previously studied in unweighted networks \cite{krl00,dms00,bb01a,bb01b}.    


Let us suppose, for example, that (i) each new vertex becomes attached to all the ends of $m\geq 1$ preferentially selected edges, (ii) the preference function is a linear function: $f(w)=w+a$ where $a>-1$, (iii) weights of the selected edges are increased by a constant $\Delta$. In this case, formulas (\ref{e5b}) and (\ref{e5c}) take the forms:  

\begin{eqnarray}
& & 
\gamma_w = 
2 + 2\,\frac{1+a}{\Delta}  
\, , 
\label{e19}
\\[5pt] 
& &
\gamma_s = \gamma_k = 
2 + \frac{1+a}{1+\Delta+a}
\, . 
\label{e20}
\end{eqnarray} 
The derivation of these expressions is very similar to that in Sec.~\ref{s-model} [one must take into account relation (\ref{e5})]. 
If  $a \to -1$, then all the three exponents $\gamma_w, \gamma_s, \gamma_k \to 2$. Note that if $a \to \infty$, then $\gamma_s, \gamma_k \to 3$. 
One can easily understand this limit. The infinite $a$ actually means the absence of preference, and edges are chosen at random, without accounting for their weights, exactly as in our network of Ref.~\cite{dms01}. So, we arrive at the same value of exponents as in that model. (Recall that $\gamma_k=3$ in the Barab\'asi-Albert model \cite{ba99}.) 



\begin{figure}
\epsfxsize=68mm
\epsffile{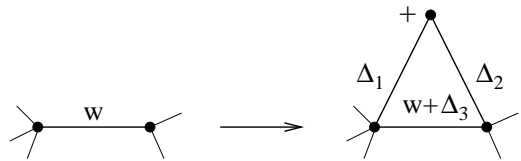}
\caption{
A generalization of the network growth process shown in Fig.~\protect\ref{f3}. 
The weights of new edges, 
$\Delta_1$ and $\Delta_2$, and the addition $\Delta_3$ may be random.  
}
\label{f6m}
\end{figure}


Another natural generalization of the process in Fig.~\ref{f3} is shown in Fig.~\ref{f6m}. In this process, new edges also can have different weights. Furthermore, all three numbers, $\Delta_1$, $\Delta_2$, and $\Delta_3$, may be random.

\section{Deterministic weighted 
networks (pseudofractals)}\label{s-deterministic}

Compact growing networks with the small-world effect may be produced in a deterministic way \cite{brv01,
dgm02,jkk02,noh03,d03,nr04,cfr04,ahas04,lgh04,gsh03}. These graphs have a discrete spectrum of degrees. These spectra may have a variety of shapes: they may be fat-tailed, in particular, scale-free, they may be rapidly decreasing, e.g., exponential. 
The key and necessary feature of these deterministic graphs is the small-world effect: their mean intervertex distances grow slower than any (positive) power of the numbers of vertices. So, while these graphs visually look very similar to fractals, they are only parodies of fractals. The difference is crucial. Fractals have a finite dimension (i.e., their intervertex distances grow as a power of the numbers of their vertices). In contrast, the networks, which we discuss, have infinite dimension (i.e., their intervertex distances grow slower than any power of the numbers of vertices, e.g., logarithmically). This is why we call deterministic graphs of this kind {\em pseudofractals} \cite{dm02,dgm02}. 

One should stress that a scale-free architecture (a power-law degree spectrum) is not not a definitive, necessary feature of these graphs. Many fractals have scale-free spectra of degrees (see discussion in Ref.~\cite{dmbook03}). 

How may these deterministic compact networks with a self-similar structure be constructed? There are two ways to obtain a compact deterministic network:

\begin{list}{}{} 

\item[(i)]
The first was realized in Ref.~\cite{brv01}. At each step of the evolution, generate a number of the copies of a graph and connect them together, e.g., by adding new edges, which decreases intervertex distances. 

\item[(ii)] 
One can use the following approach \cite{dm02,dgm02}. 
At each step, in a regular manner, transform given basic subgraphs of the network into other, larger but compact, configurations consisting of the same basic clusters. These basic clusters may be vertices, edges, triangles, squares, etc. The words ``transform in a regular manner'', in the simplest, very particular case, mean ``transform all the given basic subgraphs of the graph''. 
The words ``larger but compact configurations'' are illustrated by an example below. 

\end{list} 
(Note that in some specific situations, these approaches are equivalent.)


\begin{figure}[b]
\epsfxsize=52mm
\epsffile{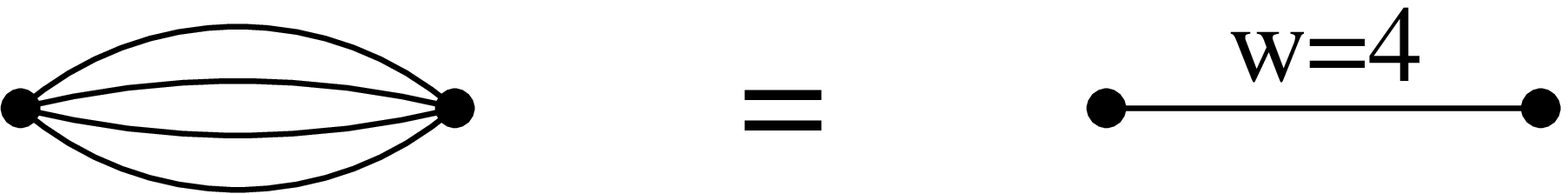}
\caption{
Multiple edges of unweighted networks may be tre\-a\-ted as single weighted edges. 
}
\label{f6}
\end{figure}



\begin{figure}
\epsfxsize=63mm
\epsffile{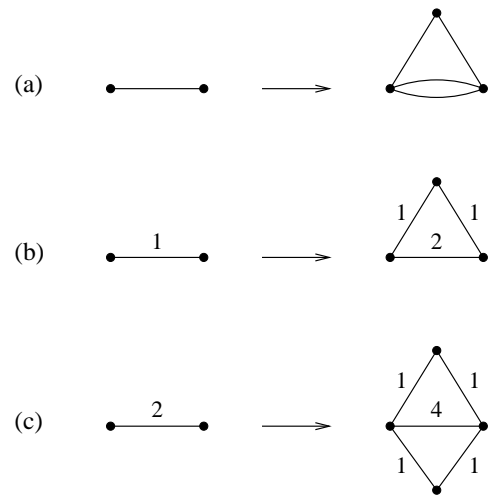}
\caption{
(a) An example of transformation of ``elementary'' edges. $\Delta=1$ 
(compare with Fig.~\protect\ref{f3}). 
(b) An equivalent representation of the transformation of an edge of weight $1$. (c) The resulting transformation of an edge of weight $2$ following from (a) or (b). 
}
\label{f7}
\end{figure}



\begin{figure}
\epsfxsize=63mm
\epsffile{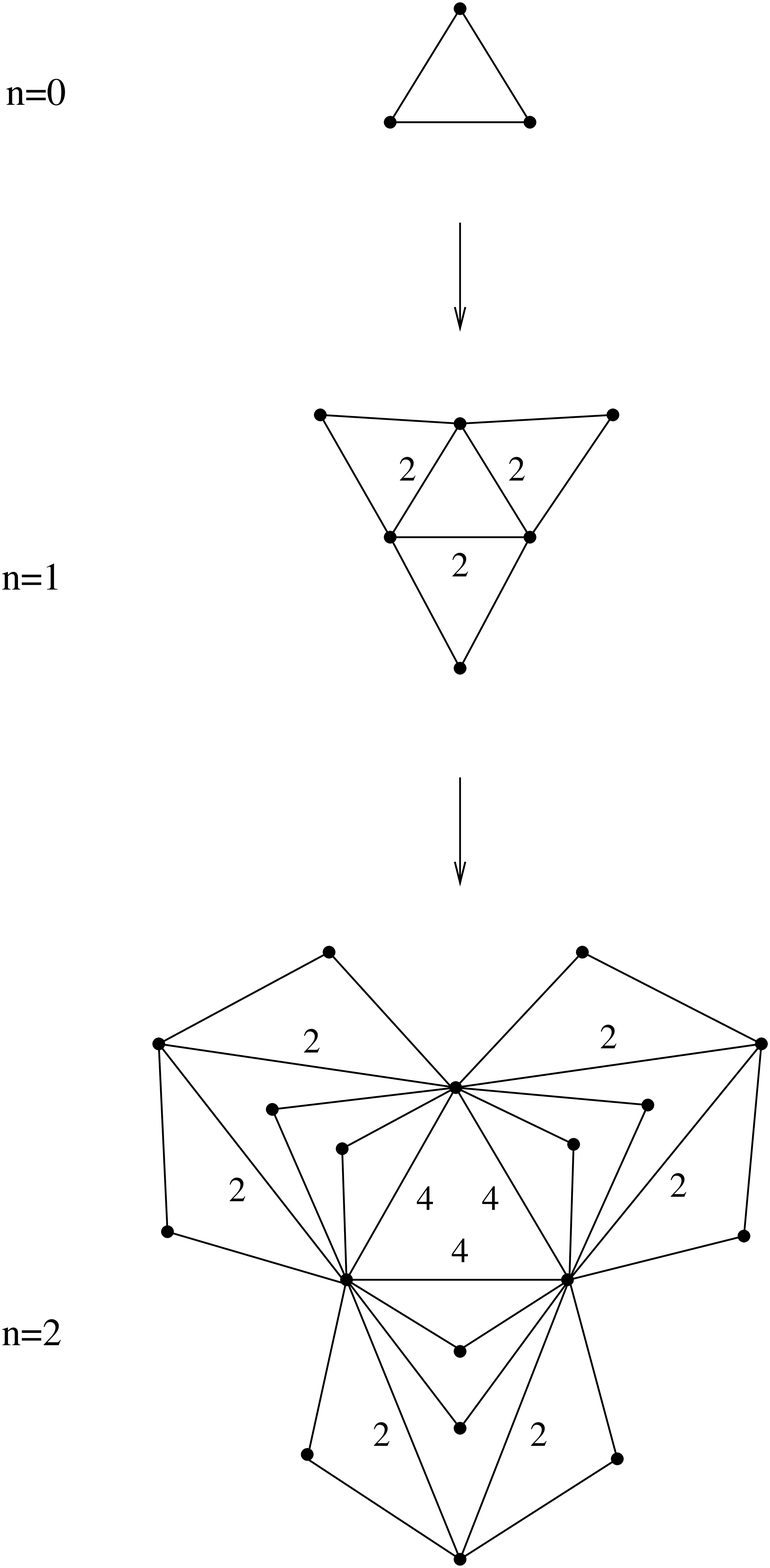}
\caption{
An example of a deterministic, scale-free, weighted graph with the small-world effect. The graph is obtained by successive application of the transformation from Fig.~\protect\ref{f7} ($\Delta=1$) to all the edges of the graph. 
The evolution starts from the triangle of edges of weight $1$ ($n=0$).  
Only the first two steps are shown. 
The bare edges denote the edges of weight $1$.
}
\label{f8}
\end{figure}


Let us demonstrate how to construct a deterministic, scale-free, weighted, compact network by using the second approach. We will exploit the equivalent representation of integer-weighted edges as multiple connections (see Fig.~\ref{f6}). 

As an example, we use the transformation that is a deterministic variation of the transformation shown in Fig.~\ref{f3} with an integer $\Delta$. This transformation is shown in Fig.~\ref{f7}. The edge of integer weight $w$ is transformed into the edge of weight $w(1+\Delta)$ with $w$ triangles of $1$-weight edges attached. 
The growth starts from the triangle of edges of weight $1$. 
At each successive step, each edge of the graph is transformed in the way shown in Fig.~\ref{f7}. The result is a weighted pseudofractal network, shown in Fig.~\ref{f8} in particular case of $\Delta=1$. If $\Delta=0$, we get the deterministic graph of Refs.~\cite{dm02,dgm02}.  

We will show below that the weight, strength, and degree distributions of this graph are power-law with exponents 

\begin{eqnarray}
& & 
\gamma_w = 1 + \frac{\ln(3+\Delta)}{\ln(1+\Delta)}
\, , 
\label{e21}
\\[5pt] 
& &
\gamma_s = \gamma_k = 1 + \frac{\ln(3+\Delta)}{\ln(2+\Delta)}
\, ,  
\label{e22}
\end{eqnarray} 
respectively. 
Note that $\gamma_w,\gamma_s,\gamma_k \to 2$ as $\Delta \to \infty$. 

We emphasize that this is only a particular example. The readers can easily consider numerous variations: use other transformations of this kind or their combinations, transform different clusters, 
transform not each of given elements of a graph but only some of them, selected in a regular way, consider networks embedded in a finite dimension space, consider trees, and so on.  
These variations produce a wide range of network architectures and a wide range of the exponents values in the range $(2,\infty)$. 

Let us consider the growth of the graph. 
One can see that at each step of the evolution, a weight of each individual edge of the graph, a strength of each individual vertex and its degree transform in the following way:  

\begin{eqnarray}
& & 
w\ \to\ w' = (1+\Delta)w
\, , 
\nonumber
\\[5pt] 
& & 
s\ \to\ s' = (2+\Delta)s
\, , 
\nonumber
\\[5pt] 
& &
k\ \to\ k' = k+s
\, , 
\label{e22a}
\end{eqnarray} 
respectively (the indices of edges and vertices are not shown). 

We introduce the following notations for basic numbers: $N_n$, $L_n$, and $W_n$, which are the total numbers of vertices and edges and the total weight of the edges in this deterministic graph in an $n$th generation, respectively. The total degree is $K_n=2L_n$. The total strength of vertices is $S_n = 2W_n$. 

In the initial state, $N_0=L_0=W_0=3$. Then, one can obtain 
 
\begin{equation}
W_n = 3(3+\Delta)^n
\, . 
\label{e23}
\end{equation} 
One can see that 

\begin{eqnarray}
& & 
N_{n+1} = N_n + W_n
\, , 
\nonumber
\\[5pt] 
& &
L_{n+1} = L_n + 2W_n
\, .  
\label{e24}
\end{eqnarray} 
This gives 

\begin{eqnarray}
& & 
N_n = \frac{3}{2+\Delta}[(3+\Delta)^n + 1+\Delta]
\, , 
\nonumber
\\[5pt] 
& &
L_n = \frac{3}{2+\Delta}[2(3+\Delta)^n + \Delta]
\, .  
\label{e25}
\end{eqnarray} 

To find the weight, strength, and degree spectra we take into account the following circumstances. 
New edges have weight $1$, new vertices have degree and strength equal to $2$. 
All the edges that emerge simultaneously have the same weight. All the vertices that emerge simultaneously have the same strength and the same degree. 
The weight $w_m$ of edges emerged $m \leq n$ generation ago and the strength $s_m$ and the degree $k_m$ of vertices emerged at that moment are 

\begin{eqnarray}
& &  
w_m = (1+\Delta)^{m-1}
\, , 
\nonumber
\\[5pt] 
& & 
s_m = 2(2+\Delta)^{m-1}
\, , 
\nonumber
\\[5pt] 
& &
k_m = \frac{2}{1+\Delta}[(2+\Delta)^{m-1}+\Delta]
\, .  
\label{e26}
\end{eqnarray} 
The $n$th generation graph contains $W_{n-1}=3(3+\Delta)^{n-1}$ new vertices and $2W_{n-1}=6(3+\Delta)^{n-1}$ new edges. 
Consequently, the $n$th generation graph contains $6(3+\Delta)^{n-m-1}$ edges of weight $w_m$ and $3(3+\Delta)^{n-m-1}$ vertices of strength $s_m$ and degree $k_m$. Here $w_m$, $s_m$, and $k_m$ are given by formula (\ref{e26}). 

As a result, we obtain a power-law behavior of the spectrum of the weight: the number of the edges of weight $w_m$ is $N(w_m) \propto w_m^{-\ln(3+\Delta)/\ln(1+\Delta)}$. This is a discrete spectrum with gaps between discrete weights growing with $w$. To present our result in a form which can be compared with results for stochastic models, where distributions are continuous and spectrum gaps are absent, we pass to the cumulative weight distribution: $\sum_{w_l\geq w_m}N(w_l) \propto w_m^{-\ln(3+\Delta)/\ln(1+\Delta)}$. This distribution already has no gaps like the corresponding cumulative distribution of the edge weight 
of stochastic networks, $N_{\text{cum}}(w) \propto w^{-(\gamma_w-1)}$. 
So we arrive at formula (\ref{e21}) for the $\gamma_w$ exponent of the edge-weight distribution. 
   
Similarly, we obtain the power strength spectrum: the number of vertices of strength $s_m$  is $N(s_m) \propto s_m^{-\ln(3+\Delta)/\ln(2+\Delta)}$. 
The cumulative distribution of the vertex strength is 
$\sum_{s_l\geq _m}N(s_l) \propto s_m^{-\ln(3+\Delta)/\ln(2+\Delta)}$, and the corresponding cumulative continuum distribution of the vertex strength in stochastic models is 
$N_{\text{cum}}(s) \propto s^{-(\gamma_s-1)}$. One can see that the degree spectrum of the deterministic graph is quite similar to the strength spectrum. So, we finally obtain formula (\ref{e22}) for the exponents of the strength and degree distributions. 

Other structural characteristics of this graph also can be easily found. 
We present here only an expression for the standard local degree-dependent clustering, which is defined as  
 
\begin{equation}
C(k)\!\equiv\! \frac
{\text{mean\! \#\! of triangles attached\! to\! a\! vertex of deg.}\,k}
{k(k-1)/2}
. 
\label{e28}
\end{equation} 
By construction, a vertex of degree $k$ in this graph has $k-1$ triangles attached, so we readily find 
 
\begin{equation}
C(k)=2/k
\,  
\label{e28}
\end{equation} 
Note that this expression is independent of $\Delta$, and we have the same result as for the network of Ref.~\cite{dm02,dgm02}.

\section{The rationale behind these models}\label{s-applications}

We discussed networks where 
edges of high weight attract new connections. 
Does this occur in the real world? 
Here we present only one illustration. 

Let us consider the evolution of a weighted network of scientific coauthorships, taken in a one-mode representation. So, the vertices are authors, and the weighted edges are (pair-wise) coauthorships. Intensive pair-wise collaborations have a greater chance to attract new collaborators than occasional connections. 
Speaking in simple terms, if two coauthors have only a single paper, e.g., written 20 years ago, there is a small chance that somebody suddenly decide to write a paper with them. In contrast, fruitful joint efforts lead to new coauthorships much more frequently. 
One may say, it is the papers---scientific results, but not their authors that attract new coauthors. This is precisely the mechanism that is discussed in the present paper.

\section{Discussion and summary}\label{s-summary}

One point should be stressed. We have shown that our approach leads to results which are close to those obtained in Refs.~\cite{bbpv04,bbv04,bbv04a}. So that, these two approaches are related. In the model of Refs.~\cite{bbpv04,bbv04,bbv04a}, ``strong'' vertices attract new connections and afterwards the weights of the edges of these vertices are specifically modified. 
In contrast, in our case, links of high weight increase their weights and attract new connections. That is, in one approach, the attachment is to (``strong'') vertices and in the other approach, the attachment is to (``heavy'') edges. This is a principal difference which is related to distinct real situations. We have shown that the introduced mechanism operates in real-world networks.   

We have demonstrated that our approach allows one to formulate minimal, non-trivial, solvable models of evolving scale-free, weighted networks. We have found a number of basic structural characteristics of these networks. We indicated some possible generalizations. In particular, we have introduced a wide class of weighted deterministic graphs of a pseudofractal type and have considered the simplest scale-free, weighted, deterministic graph in detail. 

In summary, we have developed an effective, simple approach to evolving weighted networks. This approach allows one to extend numerous results for well studied unweighted networks to a wide class of weighted \vspace{3pt}networks.   

{\em Note added.} 
After this paper had been prepared, works \cite{hjw04,hjd04,ak04}, where the linking process also is determined by the edge weights, have appeared in the cond-mat electronic archive. 


\begin{acknowledgments}
 
This work was partially supported by projects 
POCTI/FAT/46241/2002\,and\,POCTI/MAT/46176/2002. 
A part of this work was made when one of the authors (SD) attended the Exystence Thematic Institute on Networks and Risks (Collegium Budapest, June 2004). 
SD thanks A.~Barrat and A.~Vespignani for useful dis\-cussions in 
Budapest. 

\end{acknowledgments}


\end{document}